\documentclass[12pt]{article}
\usepackage{a4}
\usepackage{epsf}

\newcommand{\be}{\begin{equation}}
\newcommand{\ee}{\end{equation}}
\newcommand{\bea}{\begin{eqnarray}}
\newcommand{\eea}{\end{eqnarray}}
\newcommand{\beao}{\begin{eqnarray*}}
\newcommand{\eeao}{\end{eqnarray*}}
\newcommand{\nn}{\nonumber}
\newcommand{\pa}{\partial}
\newcommand{\e}{{\rm e}} 
 
\renewcommand{\i}{{\rm i}}
\renewcommand{\d}{{\rm d}}
\newcommand{\Tr}{{\rm Tr~}}

\newcommand{\Ref}[1]{(\ref{#1})}
\newcommand{\ep}{{\epsilon}}

\newcommand{\bc}{boundary conditions }
\newcommand{\gze}{ground state energy }
\renewcommand{\em}{electromagnetic }
\newcommand{\om}{\omega}

\begin{document}
\title{Radiative Correction to the Casimir Force on a Sphere}

\author{M Bordag and J Lindig\\
        \small Institute for Theoretical Physics, University of
               Leipzig,\\ 
        \small  Augustusplatz 10, 04109 Leipzig, Germany}

\maketitle

\begin{abstract}
The first radiative correction to the Casimir energy of a
perfectly conducting spherical shell is calculated. The 
calculation is performed in the framework of covariant perturbation 
theory with the boundary conditions implemented as constraints.
The formalism is briefly reviewed and its use is explained by 
deriving the known results for two parallel planes. 

The ultraviolet divergencies are shown to have the same structure 
as those for a massive field in zeroth order of $\alpha$. In the 
zeta--functional regularization employed by us no divergencies appear.
 
If the radius of the sphere is large compared to the Compton wavelength 
of the electron the radiative correction is of order $\alpha/(R^2m_e)$
and contains a logarithmic dependence on $m_eR$. It has the opposite sign 
but the same order of magnitude as in the case of two parallel planes.
\end{abstract}
%\pacs{}

\section{Introduction}

The Casimir effect is one of the basic effects in Quantum
Electrodynamics (QED) and provided the first example of 
{\it Nullpunktsenergie} in a field theory.  Proposed in 1948 and 
qualitatively verified in 1958, it has only recently been tested 
quantitatively \cite{lamoreaux97}. Based on the attractive force that is 
exerted on the conducting plates, Casimir even proposed a model for the 
electron. This model looked quite promising until Boyer showed a repulsive 
force in the first calculation for a sphere in 1968. In a large number 
of calculations for different geometries the Casimir force was found to 
be repulsive for configurations whose sizes in different directions are 
close to one another (a cube, for instance) and attractive when at least 
one of the sizes is much larger than the others (a long cylinder, 
for instance).  Nevertheless, a general understanding of this property is 
still missing.

Radiative corrections have rarely been considered because 
the expected effect, being proportional to the fine structure constant
$\alpha$ and at least to the ratio of the Compton wavelength $\lambda_{c}$ 
of the electron to a typical geometric size $L$ of the boundaries, is too 
small to be directly observable. Nevertheless, 
there is a general interest in their consideration, mainly having in mind 
the applicability of the general methods of quantum field theory. For 
example, its covariant formulation in the presence of explicitly 
non--covariant boundaries and 
the generalization of the perturbation expansion of a gauge theory are 
among the important issues to be addressed in this context. Boundary 
conditions necessitate a modification of the renormalization procedure as 
well. The interaction of the quantum fields with the boundary leads to
ultraviolet divergencies in vacuum graphs that cannot simply be discarded
by normal ordering. Last but not least the question for the order of the
geometrical effects is raised, i.e. the question for the leading power
of $\lambda_{c}/L$ contributing to the effective action. Additional 
attention to radiative corrections arises in the bag model of QCD where 
they are not a priori negligible.

The calculation of the ground state energy can be performed starting from 
the basic relation 
\be E_{0}={1\over 2}\sum_{(n)} E_{(n)}\;,
\label{E0}
\ee
where $E_{(n)}$ are the one particle energies of the considered system and 
the sum runs over the corresponding spectrum. This form, modified by some 
regularization (e.g., the zeta functional one with 
$E_{0}^{\rm reg}=(1/2)\sum_{(n)}E_{(n)}^{1-2s}$, ($s> 3/2$)), 
is best suited for most of the calculations for different boundary 
conditions and geometries. 
>From the point of view of general quantum field theory it is connected 
with the vacuum expectation value of the energy-momentum tensor 
$<0|\hat{T}_{\mu\nu}|0>$ by $E_{0}=\int\d\vec{x}<0|\hat{T}_{00}|0>$ 
(leaving aside modifications due to translational invariant directions). 
By means of the well known relations 
$\hat{T}_{00}={\pa \over \pa x_{\rho}}\hat{\varphi}(x){\pa \over \pa
  x_{\rho}}\hat{\varphi}(x)$ (real scalar field),
$<0|\hat{\varphi}(x)\hat{\varphi}(y)|0>=i D(x,y)$ and
$<0|\hat{T}_{00}|0>=i{\pa\over \pa x_{\rho}}{\pa\over \pa
  y_{\rho}}D(x,y)_{|_{y=x}}$ it is qualified as a one loop contribution and
can be represented as the Feynmann graph \raisebox{-6mm}{\epsffile{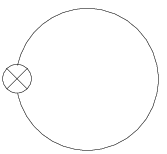}},
where \raisebox{-1mm}{\epsffile{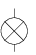}} is the vertex factor corresponding to
$T_{\mu\nu}$. Using this language it is straightforward to incorporate
radiative corrections as higher loops. For instance, in $\varphi^{4}$-theory
we would have \raisebox{-6mm}{\epsffile{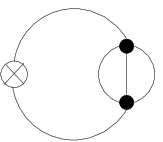}}. In QED there are two
contributions to the energy-momentum tensor, one resulting from the
electromagnetic field \raisebox{-7mm}{\epsffile{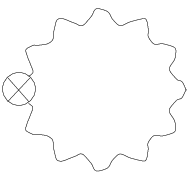}} and one from the
spinor field \raisebox{-7mm}{\epsffile{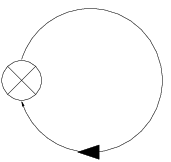}}. The corresponding two loop
contributions are \raisebox{-7mm}{\epsffile{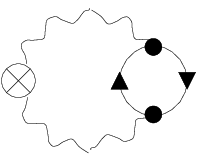}} and
\raisebox{-7mm}{\epsffile{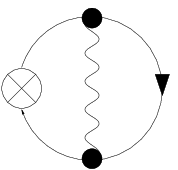}}.
 
Currently there is no established formalism for the calculation of radiative
corrections in the presence of boundaries, although a number of attempts have
been undertaken \cite{kongravndal97,xue89}. In the present paper we use the
formalism of general covariant perturbation theory, including the standard
theory of renormalization with the 'minimal' modifications needed and calculate 
the first radiative correction to the Casimir energy for a perfectly
conducting spherical shell. The basic ideas are taken from an earlier work
\cite{borowi85}, however the time passed shows that the required modifications
are not simple.

The conductor \bc  the \em field has to fulfil on some boundary ${\cal S}$ read
\be n^{\mu}F^{*}_{\mu\nu}(x)_{|_{x\in {\cal S}}}=0\;,
\label{cbc}\ee
where $F^{*}_{\mu\nu}=\epsilon_{\mu\nu\alpha\beta}F^{\alpha\beta}$ is the dual
field strength tensor and $n^{\mu}$ is the (outer) normal of ${\cal S}$. 
Being the idealization of a physical interaction (with the conductor),
they are formulated in terms of the field strengths and thus gauge--invariant. 
Now, for well known reasons, it is desirable to perform the quantization of 
QED in terms of the gauge potentials $A_{\mu}(x)$. Obviously, the \bc
\Ref{cbc} do not unambiguously imply \bc for all components of $A_{\mu}(x)$ 
as it is required in order to obtain a selfadjoint wave operator. 
 
There are two possibilities to proceed. The first one is to impose 
boundary conditions on the potentials in such a way that the conditions 
\Ref{cbc} are satisfied and a selfadjoint wave operator is provided. 
Electric and magnetic boundary conditions are commonly used for this 
purpose. These conditions are stronger than \Ref{cbc} but they are not gauge 
invariant. When now requiring BRST invariance the ghosts become boundary 
dependent too and contribute to physical quantities like the Nullpunktsenergie. 
This was first observed in the papers \cite{ambjornhughes83} and has later also 
been  noticed in connection with some models in quantum cosmology 
\cite{kamench}. The common understanding is that the ghost 
contributions cancel those from the unphysical photon polarizations, for recent 
discussions see \cite{espositokamenshchikkirsten97,BoKiVass}. 

In a second approach one considers the boundary conditions \Ref{cbc} as 
constraints when quantizing the potentials $A_{\mu}(x)$ as it was first 
done in \cite{borowi85}. In that case, explicit gauge invariance is kept. 
There is no need to impose any additional conditions. The conditions 
\Ref{cbc} appear to be incorporated in a 'minimal' manner. In that respect, 
this second approach resembles the so called dyadic formalism which has 
succesfully been used in the calculation of the Casimir energy in spherical 
geometry \cite{miltonderaadschwinger78}.

There are two ways to put this approach into practice. The first way is 
to solve the constraints explicitly. For this purpose one has to introduce 
a basis of polarization vectors $E_{\mu}^{s}$ (instead of the commonly used 
$e_{\mu}^{s}$) such that only two amplitudes (in our notation those with 
$s=1,2$) of the corresponding decomposition 
\be
A_{\mu}(x)=\sum\limits_{s=0}^{3}E_{\mu}^{s}a_{s}(x)
\label{dec1}
\ee 
of the \em potential have to satisfy boundary conditions. The other two amplitudes 
(in our notation those with $s=0,3$) remain free. In the case where the 
surface ${\cal S}$ consists of two parallel planes such a polarization basis can 
be constructed explicitly \cite{bordag84}
\be
\begin{array}{rclrcl}
E_{\mu}^{1}&=&\left(\begin{array}{c} 0\\ i\pa_{x_{2}}\\ -\i
    \pa_{x_{1}}\\0\end{array}\right){1\over\sqrt{-\pa^{2}_{x_{||}}}}, &
E_{\mu}^{2}&=&\left(\begin{array}{c} -\pa^{2}_{x_{||}}\\
    -\pa_{x_{0}}\pa_{x_{1}} \\ -\pa_{x_{0}}\pa_{x_{2}}\\0\end{array}
\right){1\over\sqrt{-\pa^{2}_{x_{||}}}\sqrt{-\pa^{2}_{x_{0}}+
    \pa^{2}_{x_{||}}}},\\
 E_{\mu}^{3}&=&\left(\begin{array}{c}0\\0\\0\\1\end{array}\right),&
E_{\mu}^{0}&=&\left(\begin{array}{c}-i\pa_{x_{0}}\\-i\pa_{x_{1}}\\-i\pa_{x_{2}}\\0\end{array}\right){1\over\sqrt{-\pa^{2}_{x_{0}}+\pa^{2}_{x_{||}}}}
\end{array}
\label{polbas1}
\ee
(${E_{\mu}^{s}}^{\dag}g^{\mu\nu}E_{\nu}^{s}=g^{st}$). 
Inserting the decomposition \Ref{dec1} with these polarization vectors into 
the \bc \Ref{cbc}, we find
\be 
a_{1}(x)=a_{2}(x)=0 ~~~~\mbox{for} ~~~~ x\in{\cal S}\; ,
\label{pampl}
\ee
whereas the amplitudes $a_{s}(x)$ with $s=0,3$ are unaffected by the 
boundary conditions. 
In this way, roughly speaking half of the photon polarizations feel the 
boundary (in \cite{bordag84a} they have been shown to be the physical ones 
in the sense of the Gupta--Bleuler quantization procedure) and half of them do
not. However, such an explicit decomposition, which simultaneously diagonalizes 
both, the action and the \bc can only be found in the simplest case. 
The problem is that for a non--planar surface ${\cal S}$ the polarizations 
$E_{\mu}^{s}$ become position dependent. To the authors' knowledge, their 
existence in the general case is still an open issue. 

Here we follow the second way to realize the boundary conditions as 
constraints, namely we use Lagrange multipliers or, equivalently, 
restrict the integration space in the functional integral approach. 
The generating functional of the Green functions in QED reads 
\bea \label{gf}
Z(J,\overline{\eta},\eta )&=& C\int{\cal D}A~{\cal
  D}\overline{\psi}~{\cal D}\psi ~
\prod_\nu\prod _{x\in \cal S}\delta \left(n^\mu F^{*}_{\mu\nu}(x)\right)
\exp\left\{ iS\left[A_\mu, \psi,\overline{\psi}\right]\phantom{\int}\right.\nn\\
% ~~~~~~~~~~~~~~~~~~~~~~~~~~~~~~
&&\left.\phantom{dd}+i\int{\rm d}x~\left(A_\mu(x)J^\mu (x)+
\overline{\psi}(x)\eta (x)+
\overline{\eta}(x)\psi (x)\right)\right\}\;,
\eea
where the integration runs over all fields with the usual asymptotic 
behavior. The functional delta function restricts the integration space to
such potentials $A_{\mu}$ that the corresponding field strengths satisfy
the \bc \Ref{cbc}. This approach had been used in \cite{borowi85} in the 
case of plane parallel boundaries. It was shown to result in a new photon 
propagator and an otherwise unaltered covariant perturbation theory of QED. 
As it will be seen in the next section, the \bc \Ref{cbc} appear to be 
incorporated with a 'minimal' disturbance of the standard formalism,
completely preserving gauge invariance (as well as the gauge fixing procedure) 
and Lorentz covariance as far as possible. An additional advantage of this 
approach is the fact that the resulting formulas for practical calculations 
turn out to be rather simple. 

The spinor field deserves a special discussion with respect to the boundary 
conditions. 
We do not impose \bc on the electron and consider the \em field and the 
spinor field on the entire Minkowski space with the conducting surface 
placed in it. In the case of ${\cal S}$ being a sphere we thus consider the 
interior and the exterior region together. In general, the surface ${\cal S}$ 
need not be closed. Only the electromagnetic field obeys boundary conditions 
on the surface ${\cal S}$. The electron penetrates it freely, it does not feel 
the surface. As for a physical model one can think of a very thin metallic 
surface which does not scatter the electrons but reflects the electromagnetic 
waves. If the thickness of the metallic surface (e.g. $1\,\mu$m) is small 
compared to the radiation length of the metal (e.g. $1.43\,$cm for copper 
\cite{PDG}), this approximation is well justified. However, the radiative 
corrections are of order $(\alpha\lambda_{c}/L)$ with respect to the Casimir 
force itself and thus too small to be directly observable. A discussion 
of the validity of this approximation seems therefore a bit academic.

We remark, that the situation is different in the case of the bag model.
The boundary conditions of the gluon field and of the spinor field are 
connected by means of the equation of motion (this is because the field 
strength tensor enters the boundary conditions rather than the dual field 
strength tensor in \Ref{cbc}). Also, we cannot a priori expect the radiative 
corrections to be very small. However, a detailed discussion 
of this subject goes beyond the scope of the present paper. Nevertheless, 
we would like to regard the investigation of the radiative corrections in 
QED as a necessary step towards the corresponding calculations in QCD.

For the needs of renormalization we consider the quantum fields embedded in a 
classical system, given by the surface ${\cal S}$ with a corresponding classical 
energy. For example, the classical energy of a sphere takes the form 
\be 
E_{0}^{\rm class}= p V + \sigma S + F R +k+\frac{h}{R}\;,\label{Eclass1}
\ee 
where $V=\frac{4}{3}\pi R^3$ and $S=4\pi R^2$ are the volume and the surface 
area, respectively. The basic idea is to renormalize the divergencies of the 
ground state energy by a redefinition of the corresponding parameters 
$p$, $\sigma$, $F$, $k$, $h$ in the classical part of the system. 
For recent comprehensive discussions we refer to the papers \cite{wipf,wir}. 
The general structure of the potentially divergent contributions  for an 
arbitrary surface ${\cal S}$ is completely known from the heat kernel expansion. 
Explicit formulas for the sphere are given in, e.g., \cite{wir95}. 
When one considers the quantum fields in the interior as well as in the exterior 
of the sphere (as we do), the divergent contributions with odd powers of the 
radius $R$ cancel. Therefore, it is sufficient to
keep only the two contributions 
\be E_{0}^{\rm class}=\sigma
S+k\;
\label{Eclass2}\ee 
in the classical energy. Hence, as a hypothesis we expect divergent 
contributions proportional to $R^{2}$ and $R^{0}$ (i.e., proportional to a 
constant) when including the radiative correction. Indeed, this turns out to 
be the case (see section 5). The corresponding redefinition of $\sigma$ and 
$k$ now depends on the fine structure constant $\alpha$ and the electron mass
$m$ (see eqs. \Ref{pol1} and \Ref{pol2} below). Besides these boundary induced 
divergencies, the ordinary free Minkowski space vacuum contributions which 
are usually discarded by normal ordering, are expected to appear as well. In 
line with the common interpretation, their removal is understood as a 
renormalization of the cosmological constant.   

It is interesting to observe the following fact. For dimensional reasons,
the divergent contributions \Ref{Eclass2} to the ground state energy are
proportional to positive powers of the mass, $\sigma \sim m^{3}$ and
$k\sim m$. Consequently, they vanish for massless fields, e.g. the pure \em 
field. However, this well known picture is not preserved when including the 
first radiative correction. Due to the interaction a dimensional parameter, 
namely the electron mass, is introduced and the divergent contributions 
proportional to $R^{2}$ and $R^{0}$ do no longer vanish. The necessity to 
consider the general strucure \Ref{Eclass2} of the classical energy in QED 
becomes evident.

In the case of plane parallel plates of separation $L$, the ground state energy 
$E_{0}$ including the radiative correction $E_0^{(1)}$ to leading order 
in $\lambda_{c}/L$ is given by \cite{borowi85} 
\be E_{0}^{\rm plates}\equiv E^{(0)}_{0}+E_0^{(1)}=-{\pi^{2}\over 720}{1\over
  L^{3}}+{\pi^{2}\over 2560}{\alpha\lambda_{c}\over
  L^{4}}\;.\label{radcorr1}
\ee 
Let us remark that the first loop correction is of order $\alpha$ and that the 
first possible contribution in the small ratio $\lambda_{c}/L$ is present. 
In fact, once this is known, the remaining task consists in the calculation 
of the number $\pi^{2}/2560$. The calculations that have appeared in the 
literature, show different results. Xua \cite{xue89} implements periodic 
boundary conditions for the potentials. In his result, all powers of 
$\lambda_{c}/L$ are absent leaving only an exponentially small contribution. 
In the paper by Kong and Ravndal \cite{kongravndal97} the same \bc as here 
are studied with the help of a different method, namely some effective field 
theory and the Euler--Heisenberg Lagrangian. Their result, 
$E_0^{(1)}\sim (\alpha^2 /L^{3})(\lambda_{c}/ L)^4$, is of one order $\alpha$
smaller than \Ref{radcorr1}. The renormalization has 
also been discussed in the paper \cite{Ford}.

In view of the result \Ref{radcorr1} the radiative correction in leading order
of $\lambda_c/R$ for a conducting sphere of radius $R$ could  be expected as 
\be 
E_{0}^{\rm sphere}={0.092353\over 2R}+\mbox{const}\cdot 
{\alpha\lambda_{c}\over R^{2}}\;,\label{radcorr2} 
\ee 
where the first term is Boyer's result (in the formulation of 
\cite{miltonderaadschwinger78}). The correction is calculated in
section 5. We find
\be
E_0^{(1)}=-7.5788\cdot 10^{-4}\,\frac{\alpha\lambda_c}{R^2}\ln(mR)
-6.4833\cdot 10^{-3}\,\frac{\alpha\lambda_c}{R^2}.
\ee
Besides the term proportional to $1/R^{2}$ it exhibits an additional 
logarithmic dependence $\ln(mR)/R^{2}$ ($m$ is the mass of the electron).

The result \Ref{radcorr1} allows for an interesting intuitive interpretation. 
Since no boundary conditions have been imposed on the spinors, a photon can 
pass the surface via the creation and subsequent annihilation of a virtual 
electron--positron pair. So the separation of the plates is effectively
enlarged due to the radiative corrections. In this sense, the result  
\Ref{radcorr1} can be viewed as a `renormalization' of the distance $L$ 
separating the plates 
\be 
L\to L\left( 1+{3\over 32}{\alpha\lambda_{c}\over L}\right)
\label{abstren}
\ee 
to leading order in $\lambda_{c}/L$. This was first noticed in \cite{borowi85}. 
As it was shown in \cite{bordag84}, the same 'distance renormalization' results 
when considering the radiative corrections to the photon states between the 
plates. This picture can also be adopted in the case of a conducting sphere. 
The radiative correction increases the effective radius $R$ according to
\be 
R\to R\left[ 1+{\alpha\lambda_{c}\over R}\left(1.4040\cdot 10^{-1} 
              +1.6413\cdot 10^{-2}\ln(mR)\right)\right]
\label{radren}
\ee 
(to leading order in $\lambda_{c}/R$).

The paper is organized as follows. In the next section we present the
procedure of quantization with constraints and in the third section we derive
the general structure of the radiative corrections. In the fourth section 
we rederive the results for parallel planes in order to explain the use
of the formalism in details. The fifth section contains the main part of the 
paper, namely the calculation of the radiative correction for spherical 
geometry. We conclude the paper with a summary and two technical appendices.  

\section{Quantization}

We start the quantization procedure with the generating functional
$Z(J,\overline{\eta},\eta)$ represented by the path integral \Ref{gf}. The
perturbation expansion is obtained by means of 
\be Z(J,\overline{\eta},\eta)= 
  \exp\left[i S_{\rm int}\left({\delta\over \delta i J},-{\delta\over\delta i\eta},
{\delta\over\delta i\overline{\eta}}\right)\right]Z^{(0)}(J,\overline{\eta},\eta)
\label{pert}
\ee 
with $S_{\rm int}(A,\overline{\psi},\psi)=\e\int{\rm d}x~\overline{\psi}(x)
\hat{A}(x)\psi (x) $ and the problem is reduced to the calculation of the 
generating functional for the non--interacting theory.

Before proceeding further we rewrite the boundary conditions \Ref{cbc} in the
following way. Let $E_{\mu}^{s}(x)$ ($s=1,2$) be the two polarization vectors in 
\Ref{dec1} with the properties
\bea 
{\pa\over\pa x_{\mu}}E_{\mu}^{s}(x)&=&0\;,\label{transv} \\
n^{\mu}E_{\mu}^{s}(x)&=&0\;\label{ort}
\eea 
for $x\in{\cal S}$ (assuming $\pa_{\mu}n_{\nu}(x)=\pa_{\nu}n_{\mu}(x)$). They span a 
space of transversal vectors  tangential to the surface ${\cal S}$. 
Note that due to \Ref{ort}, there is no derivative acting outside the
tangential space, i.e. no normal derivative in \Ref{transv}. Without loss of 
generality, we assume the normalization ${E^{s}_{\mu}}^{\dag}g^{\mu\nu}E_{\nu}^{t}=
-\delta^{st}$. We remark that the transformations 
$A_{\mu}\to A_{\mu}+\pa_{\mu}\varphi(x)$ and $A_{\mu}\to A_{\mu}+n_{\mu}\varphi(x)$
respect the \bc \Ref{cbc}, i.e. if $A_{\mu}(x)$ satisfies the boundary 
conditions, the transformed potential also does. The invariance under the first 
transformation simply says that the \bc are gauge independent. The second means 
that the projection of $A_{\mu}$ onto the normal $n^{\mu}$ of ${\cal S}$ is 
unaffected by the boundary conditions.
We therefore conclude that the boundary conditions \Ref{cbc} can equivalently be 
expressed as
\be 
E_{\mu}^sA^{\mu}(x)_{|_{x\in{\cal S}}}=0~~~~~(s=1,2)\;.\label{bc}
\ee
Explicit examples for such polarization vectors $E_{\mu}^{s}$   (besides
$E_{\mu}^{s}$ with $s=1,2$ in \Ref{polbas1} for parallel plates) are
\bea
E_{\mu}^{1}&=&\left(\begin{array}{c}i\pa_{x_{3}}\\0\\0\\i\pa_{x_{0}}
\end{array}\right){1\over\sqrt{-\pa^{2}_{x_{0}}+\pa^{2}_{x_{3}}}}\nn
\\
  E_{\mu}^{2}&=&\left(\begin{array}{c}\pa_{x_{0}}L\\-{x^{2}\over R}
(\pa^{2}_{x_{0}}-\pa^{2}_{x_{3}})\\{x^{1}\over R}(\pa^{2}_{x_{0}}-\pa^{2}_{x_{3}}) 
\\ 
\pa_{x_{3}}L\end{array}\right){1\over\sqrt{-\pa^{2}_{x_{0}}+\pa^{2}_{x_{3}}}
\sqrt{-\pa^{2}_{x_{0}}+\pa^{2}_{x_{3}}+L^{2}}}
\label{polbas2}
\eea
for a cylinder with axis along the $x_{3}$-direction, radius $R$ and the
orbital momentum operator $L=-x_{2}\pa_{x_{1}}+x_{1}\pa_{x_{2}}$, and
\bea 
E_{\mu}^{1}&=&\left(\begin{array}{c}0\\
\vec{L}\end{array}\right){1\over\sqrt{L^{2}}}\nn
  \\
E_{\mu}^{2}&=&\left(\begin{array}{c}L^{2}\\ i R\pa_{x_{0}}
(\vec{n}\times\vec{L})\end{array}\right){1\over\sqrt{L^{2}}
\sqrt{-R^{2}\pa^{2}_{x_{0}}-L^{2}}}
\label{polbas3}
\eea
for a sphere of radius $R$ with normal vector $\vec{n}=\vec{x}/|x|$ and angular 
momentum operator $\vec{L}=i\vec{x}\times\vec{\pa}_{x}$. The properties 
\Ref{transv} and \Ref{ort} as well as the normalization are easily checked. 
Evidently, these explicit polarization vectors do not contain 
normal derivatives. It should be noticed that the polarization $E^1_{\mu}$ in
the examples \Ref{polbas2} and \Ref{polbas3} coincides with that of the 
standard approach (and the dyadic approach) whereas $E^2_{\mu}$ contains
the combination $\vec{n}\times\vec{L}$ instead of $\vec{\nabla}\times\vec{L}$.

Now, with the \bc in the form \Ref{bc} we rewrite the delta functions 
in \Ref{gf} and represent them as functional Fourier integrals 
\be {
\prod_{x\in\cal S}~\delta\left( E_\mu^s(x)A^\mu (x)\right)=
\int{\cal D}b_s~\exp\left\{i\int_{z\in\cal S}{\rm d}z~
b_s(z){E_\mu^{s}}^{\dag}(z)A^\mu(z)\right\}
}\quad (s=1,2)\;,\label{frep}
\ee
where $b_s(z)$ is a field `living' on the surface $\cal S$. It corresponds to
the Lagrange multiplier in the canonical quantization approach. According to 
\Ref{pert} we turn to the calculation of the generating functional of the 
free Greens functions
\bea
Z^{(0)}(J,\overline{\eta},\eta)&=&   C\int{\rm D}A~{\cal
D}\overline{\psi}~{\cal D}\psi ~{\cal D}b~
\exp\Big\{ i\big[S^{(0)}(A)\nn \\
    &&+S^{(0)}(\overline{\psi},\psi )+
\int_{z\in\cal S}{\rm d}z~b_s(z) {E_\mu^s}^{\dag}(z) A^\mu (z)\nn \\
&&+\int {\rm d}x~\big(A_\mu (x)J^\mu (x)+
\overline{\psi}(x)\eta (x)+\overline{\eta}(x)\psi (x)\big)\big]\Big\}
\label{intb}
\eea
with $S^{(0)}(A)=\frac{1}{2}\int{\rm d}x~A_\mu(x)K^{\mu\nu}A_\nu(x)$, 
$S^{(0)}(\overline{\psi},\psi )=\int{\rm d}x~\overline{\psi}(x)\left({\rm 
i}\hat{\partial}-m\right)\psi (x)$ and $K^{\mu\nu}=g^{\mu\nu}\partial^2-(
1-1/\alpha )\partial^\mu\partial^\nu$ (hereafter, the summation 
over double latin indices from $1$ to $2$ is understood).
Since the \bc are gauge--invariant, they do not interfere with the integration
over the gauge group. Therefore, the quantization of the gauge field in covariant 
gauge with gauge parameter $\alpha$ proceeds in a standard manner, e.g., with the 
help of the Faddeev--Popov procedure. In order to diagonalize the quadratic form 
in the exponential of \Ref{intb} we substitute
\[ {
A_\mu (x)\to A_\mu (x)-\int{\rm d}y~D_{\mu\nu}(x-y)J^\nu (y) 
-\int_{z\in\cal S}{\rm d}z~D_{\mu}^{\phantom{\mu}\nu}(x-z) E^s_\nu(z)b_s(z)}
\]
where 
\be 
D_{\mu\nu}(x-y)=\left(g_{\mu\nu}-(1-\alpha)\pa_{x_{\mu}}\pa_{x_{\nu}}/\pa^{2}_{x}
\right)D(x-y)\;,\label{photprop}
\ee 
\be 
D(x-y)=\int{\d^{4}k\over (2\pi)^{4}}~{\e^{ik(x-y)}\over
    -k^{2}-i\epsilon}\quad (\ep >0)  
\label{scalprop}\ee
are the photon propagator in covariant gauge and the causal propagator,
respectively.
The result is 
\bea 
&&S^{(0)}(A)+\int_{z\in\cal S}{\rm d}z~b_s(z) {E_\mu^{s}}^{\dag}(z) A^\mu (z)+
\int {\rm d}x~A_\mu (x)J^\mu (x)\nn\\
&\to &
\frac{1}{2}\int{\rm d}x~A_\mu(x)K^{\mu\nu}A_\nu(x)-
\frac{1}{2}\int{\rm d}x\int {\rm d}y~J^\mu (x)D_{\mu\nu}(x-y)J^\nu (y)\nn\\
&&-\int{\rm d}x\int_{z\in\cal S}{\rm d}z~J^\mu(x)D_{\mu}^{\phantom{\mu}\nu}(x-z)
E_\nu^s(z)b_s(z)\nn\\
&&-\frac{1}{2}\int_{z\in\cal S}{\rm d}z\int_{z'\in\cal S}{\rm d}z'~b_s(z)
\bar{K}^{st}(z,z')b_t(z')
\nn\eea
with 
\begin{equation} {
\bar{K}^{st}(z,z')\equiv {E_\mu^{s}}^{\dag}(z)D^{\mu\nu}(z-z')E_\nu^t(z'),
~~~z,z'\in\cal S\,.
}\label{DefK}\end{equation}
The newly defined object $\bar{K}^{st}(z,z')$ is in fact the projection 
of the propagator $D_{\mu\nu}(z-z')$ on the surface $\cal S$ and, with 
respect to the Lorentz indices, into the tangential subspace spanned by the 
polarization vectors $E_\mu^s(x)$, $s=1,2$. Further we need to define the 
inversion $\mbox{$\bar{K}^{-1}$}^{st}(z,z')$ of this operation
\be 
{\int_{\cal S}{\rm d}z''~\mbox{$\bar{K}^{-1}$}^{st'}(z,z'')
\bar{K}^{t't}(z'',z')=\delta_{\cal S}(z-z')\delta_{st}
}\;,
\label{inv}
\ee
where $\delta_{\cal S}(z-z')$ is the delta function with respect to the 
integration over the surface $\cal S$, $\delta_{st}$ is the usual Kronecker 
symbol. The quadratic form in \Ref{intb} is finally diagonalized by the 
substitution 
\[ {
b_s(z)\to b_s(z)-\int_{\cal S}{\rm d}z'\int{\rm d}x~
\bar{K}^{-1}_{st}(z,z'){E_\mu^{t}}^{\dag}(z')D^\mu_{\phantom{\mu}\nu}(z'-x)
J^\nu (x)\;
}\]
and we obtain 
\bea
&&S^{(0)}(A)+\int_{z\in\cal S}{\rm d}z~b_s(z) E_\mu^s(z) A^\mu (z)+
\int {\rm d}x~A_\mu (x)J^\mu (x)\nn\\
&\to &\frac{1}{2}\int{\rm d}x~A_\mu(x)K^{\mu\nu}A_\nu(x)-
\frac{1}{2}\int_{\cal S}{\rm d}z\int_{\cal S}{\rm d}z'~b_s(z)
        \bar{K}^{st}(z,z')b_t(z')\nn\\
&&-\frac{1}{2}\int{\rm d}x\int{\rm d}y~J^\mu(x)~^SD_{\mu\nu}(x,y)J^\nu (y)
\nn
\eea
with the new photon propagator
\bea\label{bprop}
~^SD_{\mu\nu}(x,y)&\equiv & D_{\mu\nu}(x-y) - \bar{D}_{\mu\nu}(x,y)\nn
\\&=&D_{\mu\nu}(x-y)\\
&&-\int_{\cal S}{\rm d}z\int_{\cal S}{\rm d}z'~D^{\phantom{\mu}\mu '}_{\mu}(x-z)
E_{\mu '}^s(z)\bar{K}^{-1}_{st}(z,z') {E_{\nu '}^{t}}^{\dag}(z')
D^{\nu '}_{\phantom{\nu '}\nu}(z'-y).
\nn\eea
Now, the integration over all fields can be carried out and we arrive at
\bea
Z^{(0)}(J,\overline{\eta},\eta )&=&C\left(\det K\right)^{-\frac{1}{2}}
\left(\det \bar{K}\right)^{-\frac{1}{2}}\nn\\
&&\times\exp \left[ \frac{\i}{2}\int{\rm d}x\int {\rm d}y~
J^\mu (x)~^SD_{\mu\nu}(x,y) J^\nu (y) \right.\nn\\
&&+\left.\frac{1}i\int{\rm d}x\int {\rm d}y~
\overline{\psi}(x)S(x-y)\psi (y)\right]\;,
\label{Z0}
\eea
where $S(x-y)$ is the usual electron propagator. 

The generating functional \Ref{Z0} together with relation \Ref{pert} defines
the perturbation theory of QED in the conventional way. Apart from the factor 
$\left(\det \bar{K}\right)^{-\frac{1}{2}}$ due to the integration over the 
field $b_s(z)$ ($z\in\cal S$), the boundary conditions manifest themselves in 
the modified photon propagator \Ref{bprop}. Our calculation of the radiative 
corrections relies on this representation of the boundary dependent propagator.
As a consequence of \Ref{Z0} the usual language of Feynmann graphs can be 
adopted. It may be observed with the help of \Ref{transv} that the boundary 
dependent part of the propagator $\bar{D}_{\mu\nu}(x,y)$ and its part  
depending on the gauge parameter $\alpha$ are orthogonal. 

Equation \Ref{Z0} was first derived in \cite{borowi85} in a more complicated
way for the case of two parallel planes. Here it is derived for a general
surface ${\cal S}$. However, the main problem in a practical calulation consists
in finding the inverse of $\bar{K}^{st}(z,z')$. 

The representation \Ref{bprop} allows for a simple interpretation of the boundary 
dependent part $\bar{D}_{\mu\nu}(x,y)$ of the photon propagator. It can be 
viewed as describing the free propagation of a photon from the point $x$ to a point 
$z\in {\cal S}$ on the surface ${\cal S}$, a subsequent propagation on 
${\cal S}$ 
from $z$ to $z'$ mediated by $\bar{K}^{-1}(z,z')$ which is non--local in $z$ 
and $z'$, and another free propagation from $z'$ to $y$. In this way, the 
interaction of a free photon with the boundary ${\cal S}$ is accounted for, 
as symbolically indicated in fig.  1.
\begin{figure}[ht]
\centerline{\epsffile{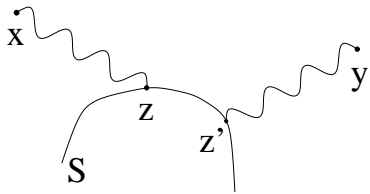}}
\label{illprop}
\caption{Illustration of the boundary dependent part
$\bar{D}_{\mu\nu}(x,y)$ of the photon propagator.}
\end{figure}

\section{Radiative Corrections}

Using the above derived  representation \Ref{Z0} we find the effective action 
by standard methods in the form
\begin{equation}{
\Gamma (A,\overline{\psi},\psi)=\frac{i}{2}{\rm Tr}\log K+
\frac{i}{2}{\rm Tr}\log \bar{K}+
S^{(0)}(A)+S^{(0)}(\overline{\psi},\psi )-i\sum \{\rm 1PI~ graphs\}\;.
}\label{effaction}
\end{equation}
The Feynman rules  are obtained from the usual ones (i.e. without boundary 
conditions) by replacing the photon propagator by its modified analogue 
\Ref{bprop}. Since the \bc are static,  the effective action is proportional to 
the total time $T$ and the ground state energy is given by
\be 
E_0=-{1\over T}\Gamma (A,\overline{\psi},\psi)_{\big | 
_{A=\overline{\psi}=\psi =0}}\;.
\label{E01}
\ee
If there are translationally invariant directions (e.g., parallel to the plates),
the relevant physical quantity is the energy density and we have to divide by the 
corresponding volume too. Graphically, the \gze is given by 1PI graphs without 
external legs. The only two--loop diagram contributing to the effective action
is \raisebox{-7mm}{\epsffile{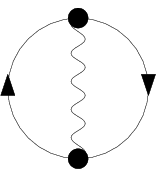}}. So up to two loop order we find the 
following expression for the ground state energy  
\bea\label{E02}
E_0\equiv E_{0}^{(0)}+E_0^{(1)}&=&-{i\over 2T}{\rm Tr}\log K 
-{i\over 2T}{\rm Tr}\log \bar{K}\\
&&+{i\over 2T}
\int {\rm d}x\int {\rm d}y~\left[D_{\nu\mu}(x-y)-
\bar{D}_{\nu\mu}(x,y)\right]\Pi^{\mu\nu}(y-x)\nn\; ,
\eea
where 
\be\label{polten}
\Pi^{\mu\nu}(x-y)=-i e^2{\rm Tr}\gamma^\mu S^c(x-y)\gamma^\nu S^c(y-x)\; 
\ee
is the known polarization tensor. The term proportional to ${\rm Tr}\log K$ 
and the contribution of the free propagator $D_{\mu\nu}(x-y)$ just constitute 
the two loop ground state energy of QED without boundary conditions. These terms 
do not depend on the geometry. According to the renormalization scheme outlined 
above, their removal can be absorbed in a redefinition of the cosmological 
constant. Therefore, they will be omitted in the following. It may be observed 
that the remaining contributions are independent of the gauge parameter 
$\alpha$, i.e. the gauge dependence of the ground state energy is completely 
contained in its free part. Let us note that the representation \Ref{E02} which 
was first used in \cite{roschawi85} is much simpler than the one previously 
applied in \cite{borowi85} where the ground state energy was expressed in terms 
of the energy-momentum tensor.

Next we insert the definiton \Ref{bprop} of $\bar{D}_{\nu\mu}(x,y)$
into the expression \Ref{E02} for the radiative correction and obtain 
(still for a general surface ${\cal S}$)
\be 
E_0^{(1)}=-{i\over 2T}\int\limits_{{\cal S}}\d z\int\limits_{{\cal S}}\d z'~
\mbox{$\bar{K}^{-1}$}^{st}(z,z')~\bar{\Pi}^{ts}(z',z)\; ,
\label{E03}
\ee
where the abbreviation
\be 
\bar{\Pi}^{ts}(z',z)=
~\int\d x\int\d y~ {E^{t}}^{\dag}_{\nu'}(z')D^{\nu'}_{\phantom{\nu}\nu}(z'-y)
 \Pi^{\nu\mu}(y-x)D_\mu^{\phantom{\mu}\mu'}(x-z)E_{\mu'}^{s}(z)_{|_{z,z'\in{\cal S}}}\;
\label{pist}
\ee
has been introduced. Using the transversality of $\Pi_{\mu\nu}(x)$,
\[ 
\Pi_{\mu\nu}(x)=(g_{\mu\nu}\pa^{2}_{x}-\pa_{x_{\mu}}\pa_{x_{\nu}})
{\Pi}(x^{2})\; ,
\]
this expression simplifies to
\[
{\bar{\Pi}}^{ts}(z',z)=\int\d x\int\d y~{E^{t}}_{\mu}^{\dag}(z')
D(z'-y)g^{\mu\nu}\pa_{y}^{2}\Pi((y-x)^{2})D(x-z)E^s_{\nu}(z)_{|_{z,z'\in{\cal S}}}\;.
\]
Introducing the Fourier transform of $\Pi(x)$
\be 
{\Pi}(x^{2})=\int{\d^{4}q\over (2\pi)^{4}}~\e^{iqx}~\tilde{\Pi}
(q^{2})
\label{Pq}
\ee
we rewrite it in the final form 
\be 
{\bar{\Pi}}^{ts}(z',z)={E_\mu^{t}}^{\dag}(z')\int{\d^{4}q\over (2\pi)^{4}}
\e^{iq(z-z')}g^{\mu\nu}{\tilde{\Pi}
(q^2)\over -q^2}E_\nu^s(z')_{|_{z,z'\in{\cal S}}}\; ,
\label{Pz}
\ee
which is best suited for the calculation of the radiative correction in
particular geometries. There is yet another important conclusion to be drawn 
from the representation \Ref{Pz}. The scalar part $\tilde{\Pi}
(q^2)$ of the polarization 
tensor is known to possess a logarithmic divergence which is independent of 
$q$ (for example, in Pauli--Villars regularization it is of the form 
$-(2\alpha/3\pi)\log(M/m)$ where $M$ is the regularizing mass). However, if only
a $q$--independent constant is added to $\tilde{\Pi}
$, i.e. $\tilde{\Pi}
(q^2)\to \tilde{\Pi}
(q^2)+c$,
the quantity ${\bar{\Pi}}^{ts}$ changes according to 
${\bar{\Pi}}^{ts}(z',z)\to {\bar{\Pi}}^{ts}(z',z)+c~\bar{K}^{ts}(z',z)$.
Then it may be verified with the help of \Ref{inv} that the corresponding
change in the ground state energy is a simple constant which is independent 
of the geometry. Consequently, the removal of the divergence of the polarization 
tensor can be interpreted as a renormalization of the cosmological constant
in complete analogy to the free Minkowski space contributions discussed above.
As a result only the finite, renormalized part of the polarization tensor 
needs to be taken into account when calculating the boundary dependent part
of the radiative correction $E_0^{(1)}$.   

\section{The ground state energy in plane geometry}

In this section we wish to demonstrate the calculational techniques needed
in the spherical case on a simple example first. For this purpose we rederive 
the results for the ground state energy $E_{0}^{(0)}$ and the radiative 
correction $E_0^{(1)}$ in the geometry of two parallel conducting planes. 
In this case, the surface ${\cal S}$ consists of two pieces. A coordinization of 
the planes is given by $z=\{x_{\alpha}, x_{3}=a_{i}\}$, where the subscript 
$i=1,2$ distinguishes the two planes and $\alpha =0,1,2$ labels the 
directions parallel to them (they are taken perpendicular to the $x_{3}$-axis 
intersecting them at $x_{3}=a_{i}$, $|a_{1}-a_{2}|\equiv L$ is the distance 
between them). The polarizations $E_{\mu}^{s}$ ($s=1,2$) are given by 
\Ref{polbas1} and do not depend on $x_{\alpha}$ or $i$. Therefore they commute 
with the free photon propagator \Ref{photprop}. Inserting them into \Ref{DefK} 
yields the operator $\bar{K}^{st}$ in the form
\be 
\bar{K}^{st}(z,z')=-\delta_{st}~D(x-x')_{|_{x,x'\in{\cal S}}}\;.
\label{ksplane}
\ee
We proceed with deriving a special representation of the scalar propagator. 
It is obtained by performing the integration over $k_{3}$ in eq. \Ref{scalprop}
\be D(x-x')=\int{\d^{3}k_{\alpha}\over
  (2\pi)^{3}}~{\e^{ik_{\alpha}(x^{\alpha}-{x'}^{\alpha})+i\Gamma
    |x^{3}-{x'}^{3}|}\over -2i\Gamma},\label{ksplane2}\ee
with $\Gamma=\sqrt{k_{0}^{2}-k_{1}^{2}-k_{2}^{2}+i\epsilon}$. Substituting 
$x_{3}=a_{i}$ and $x'_{3}=a_{j}$ we get
\be
\bar{K}^{st}(z,z')=-\delta_{st}\int{\d^{3}k_{\alpha}\over
  (2\pi)^{3}}~{i\over 2\Gamma}~h_{ij}~\e^{ik_{\alpha}(x^{\alpha}-{x'}^{\alpha})}
\label{kplane}
\ee
where the abbreviation 
\be 
h_{ij}=\e^{i\Gamma |a_{i}-a_{j}|}~~~~~~~(i,j=1,2)\;
\label{h}
\ee
has been introduced. With \Ref{kplane} we have achieved a mode decomposition 
of the operator $\bar{K}^{st}$ on the surface ${\cal S}$. As an advantage of 
this representation the inversion of $\bar{K}^{st}$, defined by \Ref{inv}, 
is now reduced to the algebraic problem of inverting the (2x2)--matrix 
$h_{ij}$. The inverse of $h_{ij}$ is given by 
\be 
h^{-1}_{ij}={i\over 2\sin\Gamma L}\left(
\begin{array}{cc}\e^{-i\Gamma
      L}&-1\\-1&\e^{-i\Gamma L}\end{array}\right)_{ij}
\label{invh}
\ee
and we get
\be 
\mbox{$\bar{K}^{-1}$}^{st}(z,z')=-\delta_{st}\int{\d^{3}k_{\alpha}\over
  (2\pi)^{3}}~{2\Gamma\over\i}~h^{-1}_{ij}~\e^{ik_{\alpha}(x^{\alpha}-{x'}^{\alpha})}\;.
\label{kiplane}
\ee
After inserting this expression into \Ref{bprop} we find the photon propagator
for the \em field in covariant gauge with conductor \bc on two parallel planes as 
proposed in \cite{borowi85}.

Next we substitute \Ref{kplane} into \Ref{E02} and obtain for the 
ground state energy density per unit area
\be 
E_{0}^{(0)}=-{i\over 2TV_{||}}\int \d^{3}x_{\alpha}\int{\d^{3}k_{\alpha}\over
  (2\pi)^{3}}\Tr\log\left(-\delta_{st}{i\over 2\Gamma}h_{ij}\right)\;,
\ee
where we have divided by the volume $V_{||}$ of the translational invariant
directions parallel to the planes. We perform the Wick rotation $k_{0}\to
ik_{0}$ (thereby $\Gamma\to i\gamma\equiv i\sqrt{k_{0}^{2}+k_{1}^{2}+k_{2}^{2}}$)
and calculate the $'\Tr\log'$ 
\[
\Tr\log\left(-\delta_{st}{1\over 2\gamma}h_{ij}\right)=
\log\det\left(-\delta_{st}{1\over 2\gamma}h_{ij}\right)=
2\log\left(1-\e^{-2\gamma L}\right)-4\log(2\gamma) \;.
\]
The term $-4\log(2\gamma)$ yields a distance independent contribution
and will be dropped. In this way the known result for the Casimir energy is
reproduced
\[ 
E_{0}^{(0)}=\int{\d^{3}k_{\alpha}\over
  (2\pi)^{3}}\log \left(1-\e^{-2\gamma L}\right)=-{\pi^{2}\over 720 L^{3}}\;.
\]

Now we turn to the radiative correction. The polarization vectors $E_{\mu}^{s}$
($s=1,2$)  commute with $\Pi$
and we find
\be
{\bar{\Pi}}^{st}(z-z')=\delta_{st}~\int{\d^{4}q\over (2\pi)^{4}}
\e^{iq(z-z')}{\tilde{\Pi}
(q^2)\over q^2}_{|_{z,z'\in{\cal S}}}\;.
\label{er2}
\ee
We substitute this expression into the radiative correction \Ref{E03} and
get 
\be 
E_0^{(1)}=\sum\limits_{i,j=1}^{2}\int{\d^{4}k\over
    (2\pi)^{4}}~\Gamma~h_{ij}^{-1}~\e^{-ik_{3}(a_{j}-a_{i})}~
{\tilde{\Pi}
(k^{2})\over k^{2}}
\ee
with $\Gamma=\sqrt{k_{\alpha}k^{\alpha}+i\epsilon}$ and
$k^{2}=k_{\mu}k^{\mu}$. 
Using \Ref{invh} the last equation takes the form
\be 
E_0^{(1)}=2i\int{\d^{4}k\over (2\pi)^{4}}~
{\Gamma\over\sin\Gamma L}~\left(\e^{-i\Gamma L}-\cos k_{3}L\right)~
{\tilde{\Pi}
(k^{2})\over k^{2}}\;.
\ee
By means of the trivial relation $\exp(-i\Gamma L)=\exp(i\Gamma L)-2i\sin\Gamma L$ 
we separate again a distance independent contribution that will be omitted.
Now we perform the Wick rotation and obtain the final result for the 
radiative correction to the ground state energy to order $\alpha$ in the 
geometry of two parallel conducting planes
\be  
E_0^{(1)}=2\int{\d^{4}k\over (2\pi)^{4}} {\gamma\over\sinh \gamma L}
\left(\e^{-\gamma L}-\cos k_{3}L\right) {\tilde{\Pi}
(k^{2})\over k^{2}}\;.
\label{radcorr}
\ee
It is interesting to study the radiative correction \Ref{radcorr} 
in the limit $\lambda_{c}/L<<1$. For this purpose we transform the 
integration path of the $k_3$--integration. Due to $\tilde{\Pi}
(k_3^2+\gamma^2)$ there is a cut
with branch point $k_{3}=i\sqrt{4m^{2}+\gamma^{2}}$ in the upper half of the
complex $k_3$--plane. The discontinuity of the one loop vacuum polarization 
$\tilde{\Pi}
(k^2)$ across the cut, 
${\rm disc}\tilde{\Pi}
(k^{2})=\tilde{\Pi}
(k^{2}+i\epsilon)
-\tilde{\Pi}
(k^{2}-i\epsilon)$, is well known
\be  
{\rm disc}\tilde{\Pi}
(k^{2})=-\frac{2i}{3}\alpha
\sqrt{1-\frac{4m^{2}}{k^{2}}}\left(1+\frac{2m^{2}}{k^{2}}\right)\;.
\label{disc}
\ee
As shown in fig. 2 we move the integration contour towards the imaginary axis so
that the cut is enclosed. The result can be written in the form
\be  
E_0^{(1)}=\frac{i}{\pi L^3}\int{\d^{3}k\over (2\pi)^{3}}
{\gamma\over\sinh \gamma L}
\int\limits_{1}^{\infty}{\d k_{3}\over k_3}
{{\rm disc}\tilde{\Pi}
\left(\frac{k_3^2}{4m^2}\right) \over \sqrt{4m^2L^2k_3^2+\gamma^2}}
\left(\e^{-\gamma}-\e^{-\sqrt{4m^2L^2k_3^2+\gamma^2}}\right)\;.
\ee
\begin{figure}[ht]
\unitlength1cm\begin{picture}(14,4)
\put(9.7,0.6){\mbox{$\Re k$}}
\put(6,3.3){\mbox{$\Im k$}}
\put(7,1.5){\mbox{$\i\sqrt{4m^{2}+q^{2}}$}}
\put(6.2,2.3){\mbox{$\gamma$}}
\centerline{\epsffile{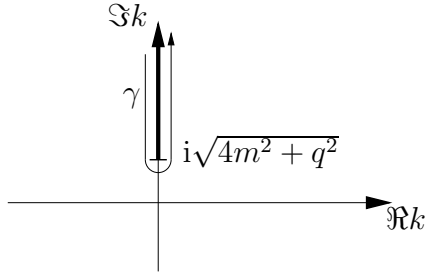}}\label{k3plane}
\end{picture}
\caption{Integration path in the $k_{3}$-plane}
\end{figure}
%\begin{figure}[ht]
%\centerline{\epsffile{Abb2.eps}}\label{k3plane}
%\caption{Integration path in the $k_{3}$-plane}
%\end{figure}
In the limit $mL>>1$, the second term in the round brackets is exponentially  
suppressed and can be neglected. The desired series in inverse powers of $mL$
is now simply achieved by expanding the square root in the denominator
\be 
E_0^{(1)}=\frac{i}{2\pi mL^4}\int{\d^{3}k\over (2\pi)^{3}}~
{\gamma~\e^{-\gamma L}\over\sinh \gamma L}~
\int\limits_{1}^{\infty}\d k_{3}
{{\rm disc}\tilde{\Pi}
\left(\frac{k_3^2}{4m^2}\right) \over k_{3}^{2}}
\left(1+O\left(\frac{\gamma^2}{mLk_3^2}\right)\right)\;.
\label{rkp2}
\ee
After some elementary integrations and with the help of 
\be 
\int\limits_{1}^{\infty}\d k_{3}
{{\rm disc}\tilde{\Pi}
\left(\frac{k_3^2}{4m^2}\right) \over k_{3}^{2}}= 
-\frac{3i\pi}{16}\alpha
\label{discint}
\ee
we find the leading order contribution to the radiative correction
\be  
\mbox{$E_0^{(1)}$}=\frac{\pi^2\alpha}{2560mL^4}+O\left(\frac{1}{m^2L^5}\right)
\label{rkorrp1}
\ee
in agreement with \cite{borowi85} and \cite{roschawi85}.

\section{The radiative correction in spherical geometry}

This section is devoted to the calculation of the radiative correction
to the ground state energy for a perfectly conducting spherical shell
of radius $R$. Appropriate coordinates $z \in {\cal S}$ are given by
$z=\{z_0, r_z=R,\theta,\phi\}$ where $r_z,\theta,\phi$ are the usual 
spherical coordinates. In order to calculate the radiative correction 
according to the representation \Ref{E03} we need to know 
$\bar{K}^{st}(z,z')$ and $\bar{\Pi}^{st}(z,z')$, i.e. the projections 
of the scalar propagator $D(x-y)$ and its product with the vacuum 
polarization, $(D\Pi)(x-y)\equiv \int\d x'~D(x-x')\Pi(x'-y)$, with respect 
to the polarization vectors $E^s_{\mu}$ of \Ref{polbas3}.

Due to the rotational symmetry it proves convenient to express these 
quantities in terms of spherical waves. Using the expansion of a plane 
wave in spherical waves
\[ 
\e^{iqx}~=4\pi\sum\limits_{lm}i^{l}j_{l}(qr_{z})Y^{*}_{lm}(\theta_{q},\phi_{q})Y_{lm}(\theta_{z},\phi_{z})\;,
\]
the scalar propagator \Ref{scalprop} can be written as
\be 
D(x-y)=\sum\limits_{lm}\int\limits_{-\infty}^{\infty}
{\d \omega\over 2\pi}\e^{i\omega(x_0-y_0)}
Y_{lm}(\theta_x,\phi_x)d_l(\omega;r_x,r_y)
Y_{lm}^*(\theta_y,\phi_y)
\label{skalpropsp}
\ee
with 
\be 
d_l(\omega;r_x,r_y)=i|\omega| j_l(|\omega| r_<)h^{(1)}_l(|\omega| r_>)\;,
\label{spprop}
\ee 
where $j_l$ and $h^{(1)}_l$ are the spherical Bessel functions and 
$Y_{lm}(\theta,\phi)$ are the surface harmonics. $r_<(r_>)$ denotes the 
smaller (larger) one of the radii $r_x$ and $r_y$. 
The corresponding spherical expansion of $(D\Pi)(x-y)$ is obtained from 
\Ref{skalpropsp} by replacing the quantity $d_l(\omega;r_x,r_y)$ by
$\Pi_l(\omega;r_x,r_y)$ whose explicit form is 
\bea
\Pi_l(\omega;r_x,r_y)&=&-\frac{2}{\pi}\int^{\infty}_0\d q q^2 
j_l(qr_x)j_l(qr_y)\frac{\tilde{\Pi}
\left(\omega^2-q^2\right)}{\omega^2-q^2}\nn\\
&=&\frac{i}{\sqrt{r_xr_y}}\int_{4m^2}^{\infty}\frac{\d q^2}{2\pi q^2}
~{\rm disc}\tilde{\Pi}
\left(q^2\right)I_{l+\frac{1}{2}}\left(pr_<\right)
K_{l+\frac{1}{2}}\left(pr_>\right)
\label{pil}
\eea
with $p=\sqrt{q^2-\omega^2}$. $I_{l+\frac{1}{2}}$, $K_{l+\frac{1}{2}}$
are the modified Bessel functions.

The polarization vector $E^{1}_{\mu}$ is proportional to the angular
momentum operator $\vec{L}$ (it corresponds to the TM--mode). It therefore
commutes with $D(z-z')$ and $(D\Pi)(z-z')$. Consequently, 
$\bar{K}^{st}(z,z')$ and $\bar{\Pi}^{st}(z,z')$ are diagonal in $s,t$ and 
their $11$--components are given by 
\be
\bar{K}^{11}(z,z')=-\sum\limits_{l\ge 1, m}
\int\limits_{-\infty}^{\infty}{\d \omega\over 2\pi}
~\e^{i\omega(z_0-{z'}_0)}
Y_{lm}(\theta_z,\phi_z)d_l(\omega;R,R)
Y_{lm}^*(\theta_{z'},\phi_{z'})
\ee
and
\be 
\bar{\Pi}^{11}(z,z')=-\sum\limits_{l\ge 1, m}
\int\limits_{-\infty}^{\infty}{\d \omega\over 2\pi}
~\e^{i\omega(z_0-{z'}_0)}
Y_{lm}(\theta_z,\phi_z)\Pi_l(\omega;R,R)
Y_{lm}^*(\theta_{z'},\phi_{z'})\; ,
\ee
respectively. Due to the projection with the angular momentum operator 
$\vec{L}$ the contribution with $l=0$ is absent.

The second polarization vector $E^{2}_{\mu}$ contains the operator
$\vec{n}\times\vec{L}$ (it is analogous but not identical to the
TE--mode). This operator does not commute with either $D(z-z')$ or 
$(D\Pi)(z-z')$. Substituting the polarization vector $E^{2}_{\mu}$ of 
\Ref{polbas3} into \Ref{DefK} we find 
\bea
\bar{K}^{22}(z,z')&=&{E^{\dag}}^{2}_{\mu}(z)g^{\mu\nu}D(z-z')E\mu^2(z')\nn\\
&=&\sum\limits_{l\ge 1,m}\int\limits_{-\infty}^{\infty}{\d \omega\over 2\pi}
~{\e^{i\omega(z_0-z'_0)}\over l(l+1)[\omega^2 R^2-l(l+1)]}~d_l(\omega;R,R)\nn\\
&&\phantom{dd}\times\left[l^2(l+1)^2Y_{lm}(\theta_z,\phi_z)
Y^*_{lm}(\theta_{z'},\phi_{z'})\right.\\
&&\phantom{dd\times}\left.-\omega^2R^2(\vec{n}_{z}\times\vec{L}_{z})^\dag 
Y_{lm}(\theta_z,\phi_z) 
Y_{lm}^*(\theta_{z'},\phi_{z'})(\vec{n}_{z'}\times\vec{L}_{z'})\right]\; .\nn
\eea
The method of reduced matrix elements which is known from elementary
quantum mechanics \cite{Landau} provides a plain way to simplify
the last expression. After a straightforward computation we arrive at
\be 
\bar{K}^{22}(z,z')=\sum\limits_{l\ge 1,m}\int\limits_{-\infty}^{\infty}
{\d \omega\over 2\pi}~\e^{i\omega(z_0-z'_0)}Y_{lm}(\theta_z,\phi_z)
f_l(\omega,R,R) Y_{lm}^*(\theta_{z'},\phi_{z'}) 
\ee
with
\be f_l(\omega;R,R)={l(l+1)d_l(\omega;R,R)-{\omega^2R^2\over 2l+1}
\left[(l+1)d_{l-1}+l d_{l+1}\right]
\over \omega^2R^2-l(l+1)}\;.
\label{fl}
\ee
Again, the corresponding result for $\bar{\Pi}^{22}(z,z')$ is obtained
from $\bar{K}^{22}(z,z')$ by replacing
$d_l(\omega;R,R)\rightarrow \Pi_l(\omega;R,R)$. Since $\bar{K}^{st}(z,z')$
is diagonal in $s,t$ as well as in the quantum numbers $\{\omega,l,m\}$
its inverse is readily found. We get
\bea 
\mbox{$\bar{K}^{-1}$}^{11}(z,z')&=&\sum\limits_{l\ge 1,m}
-\int\limits_{-\infty}^{\infty}{\d \omega\over
  2\pi R^4}~\e^{i\omega(z_0-z'_0)}
Y_{lm}(\theta_z,\phi_z){1 \over d_l(\omega;R,R)} 
Y_{lm}^*(\theta_{z'},\phi_{z'})\;,\nn \\
\mbox{$\bar{K}^{-1}$}^{22}(z,z')&=&\sum\limits_{l\ge 1,m}
\int\limits_{-\infty}^{\infty}{\d \omega\over
  2\pi R^4}~\e^{i\omega(z_0-z'_0)}
Y_{lm}(\theta_z,\phi_z){1\over f_l(\omega;R,R)}
Y_{lm}^*(\theta_{z'},\phi_{z'})\;.
\eea
Inserting these expressions into \Ref{bprop} we find the photon propagator 
in covariant gauge with conductor boundary conditions  on a sphere. We note 
that it has a somewhat unusual form because it seems not to contain radial
derivatives (corresponding to Neumann boundary conditions). However, such 
derivatives are implicitly present in the terms proportional to 
$d_{l\pm1}$ in $f_{l}$ \Ref{fl}. Utilizing the recursion relations for the 
spherical Bessel functions, $f_{l}$ can be cast into the form
\[ 
f_{l}(\omega;R,R)=-ik{(kR j_{l}(kR))'
(kR h^{(1)}_{l}(kR))'\over \om^2R^{2}-l(l+1)}\; ,\quad (k=|\omega|)\;.
\]
In fact, substituting $\bar{K}^{st}(z,z')$ with the last expression
into \Ref{E02} and performing the Wick rotation $\omega \rightarrow iy$ 
we obtain 
\be
E_0=\frac{1}{2R}\sum\limits^{\infty}_{l=1}(2l+1)\int\limits^{\infty}_{-\infty}
{\d y \over 2\pi}~\log\left(1-\lambda^2_l(x)\right)
\ee
with $x=|y|$ and 
$\lambda_l=\left(s_le_l\right)'=\left(xI_{l+\frac{1}{2}}(x)
K_{l+\frac{1}{2}}(x)\right)'$.
This mode sum representation was first derived in \cite{miltonderaadschwinger78}
with the help of the dyadic formalism. We therefore reproduce the known result 
for the Casimir energy of a conducting sphere.

Now we return to the calculation of the radiative correction. Substituting
$\mbox{$\bar{K}^{-1}$}^{st}(z,z')$ and $\bar{\Pi}^{st}(z,z')$ into
$E_0^{(1)}$ \Ref{E03} and making use of the orthogonality of the surface
harmonics one finds
\bea\label{E04}
E_0^{(1)}&=&
-\frac{i}{2}\sum\limits_{l\ge 1,m}\int{\d \omega\over 2\pi}
\left[\left(l+\frac{1}{2}\right)^2-\omega^2R^2\right]^{-\frac{\epsilon}{2}}\\
&&\times\left\{\frac{\Pi_l(\omega;R,R)}{d_l(\omega;R,R)}+\frac{l(l+1)\Pi_l
-\frac{\omega^2R^2}{2l+1}\left[l\Pi_{l+1}+(l+1)\Pi_{l-1}\right]}{l(l+1)d_l
-\frac{\omega^2R^2}{2l+1}\left[ld_{l+1}+(l+1)d_{l-1}\right]}\right\}\;,\nn
\eea
where we have introduced a regularizing factor which is removed by taking 
the limit $\epsilon\rightarrow 0$ in the end. Next we perform the Wick 
rotation $\omega\to i k$. Thereby we have from \Ref{spprop}
\be 
d_{l}(i k;R,R)=\frac{1}{R}I_{l+\frac{1}{2}}(|k|R)K_{l+\frac{1}{2}}(|k|R)
\label{dWR}
\ee
and from \Ref{pil}
\be
\Pi_l(ik;R,R)=\frac{i}{R}\int\limits_{4m^2}^{\infty}{\d q^2 \over 2\pi q^2}~
{\rm disc}\tilde{\Pi}
(q^2)~I_{l+\frac{1}{2}}(pR)K_{l+\frac{1}{2}}(pR)
\label{PWR}
\ee
with $p=\sqrt{q^2+k^2}$. Finally, we substitute $d_l$ \Ref{dWR} and $\Pi_l$ 
\Ref{PWR} into the radiative correction \Ref{E04} and obtain
\be 
E_0^{(1)}={i\over 2\pi R} \sum\limits_{l=1}^{\infty}
\int\limits_{0}^{\infty}{\d k\over \pi}~
{\nu\over \left(k^{2}+\nu^{2}\right)^{\frac{\epsilon}{2}}}
\int\limits_{4m^2R^2}^{\infty}{\d q^{2}\over q^{2}}~
{\rm disc}\tilde{\Pi}
\left(\frac{q^2}{R^2}\right)~(B_{1}+B_{2})\; ,
\label{E05}
\ee
where the abbreviations $\nu\equiv l+\frac{1}{2}$, 
\be 
B_{1}={I_{\nu}(p)K_{\nu}(p) \over I_{\nu}(k)K_{\nu}(k)}
\label{B1}
\ee
and
\be B_{2}={I_{\nu} (p)K_{\nu} (p) +{k^2 \over 2 \nu}
   \Big[\frac{1}{\nu+\frac{1}{2}}I_{\nu+1}(p)K_{\nu+1}(p)
+\frac{1}{\nu-\frac{1}{2}}I_{\nu-1}(p)K_{\nu-1}(p)\Big]
\over 
I_{\nu}(k)K_{\nu}(k)+{k^2\over 2 \nu} \Big[\frac{1}{\nu+\frac{1}{2}}
I_{\nu+1}(k)K_{\nu+1}(k)+\frac{1}{\nu-\frac{1}{2}}
I_{\nu-1}(k)K_{\nu-1}(k)\Big]},
\label{B2}
\ee
have been introduced.

Evidently, the expression \Ref{E05} for the radiative correction to the 
ground state energy of a conducting sphere contains ultraviolet 
divergencies. Hence, it needs to be renormalized. In order to isolate
the ultraviolet divergent terms we employ the uniform asymptotic expansion
of the Bessel functions for $\nu \to\infty$ and $k\to\infty$ with fixed
\be\label{st}
s\equiv\left(1+{k^{2}+q^{2}\over \nu^{2}}\right)^{-\frac{1}{2}}\;,\qquad
t\equiv\left(1+{k^{2}\over \nu^{2}}\right)^{-\frac{1}{2}}\; .
\ee
The corresponding asymptotic expansion of the quantities $B_1$ and $B_2$ 
reads 
\be\label{Bas}
B_{1,2}={s\over t}\sum\limits_{n=0}^{\infty}
\sum\limits_{i,j\ge 0 \atop n\le i,j\le 3n}
X^{(1,2)}_{nij}{s^{2i}t^{2j}\over \nu^{2n}}\; ,
\ee
where only even powers contribute. The coefficients $X^{(1,2)}_{nij}$ are 
listed in Appendix A.

Now we insert the expansions \Ref{Bas} into \Ref{E05} and perform the 
integration over $k$: 
\bea
\int\limits_{-\infty}^{\infty} \d k ~t^{\ep-1+2j}s^{1+2i}&=&
{\nu\Gamma ({\ep-1\over 2}+i+j)\Gamma ({1\over 2})\over \Gamma ({\ep\over
    2}+i+j)}\nn\\
&&\times~ {_2F}_1\left({1\over 2}+i,{\ep -1\over 2}+i+j,
{\ep \over 2}+i+j;-{q^{2}\over \nu^{2}}\right)\nn\; .
\eea
In the next step we utilize a Mellin--Barnes representation of the
hypergeometric function and rewrite the last expression as
\bea
\int\limits_{-\infty}^{\infty} \d k ~t^{\ep-1+2j}s^{1+2i}&=&
{\nu\Gamma ({1\over 2})\over \Gamma ({1\over 2}+i)}\\
&&\times \int\limits _{c}{\d\sigma\over 2\pi \i}~
{\Gamma ({1\over 2}+i+\sigma)\Gamma
  ({\ep -1\over 2}+i+j+\sigma)\Gamma (-\sigma)  \over
\Gamma ({\ep\over 2}+i+j+\sigma)}~\left({q^{2}\over \nu^{2}}\right)^{\sigma},
\nn
\label{intst}
\eea
where the integration contour $c$ goes from $-i\infty$ to $i\infty$ parallel 
to the imaginary axis with $-{1\over 2} < \Re \sigma < 0$. The sum over the 
orbital momentum $l$ can now be carried out. The result is a Hurwitz zeta
function: 
\be 
\sum\limits_{l=1}^{\infty}~\nu^{2-\ep-2n-2\sigma}=\zeta_H
(2n+2\sigma+\ep-2;{3\over 2})\;.
\label{zeta}
\ee
Here and in the previous steps we have assumed $\Re \ep >3$ so that
$\Re\ep +2\Re\sigma +2n-2 > 1$ is satisfied.
Finally, we put \Ref{disc} into the radiative correction \Ref{E05} where
we are able to perform the $q$--integration as well 
\be 
\int\limits_{1}^{\infty}{\d q^{2}\over q^{2}}~q^{2\sigma}~\sqrt{1-{1\over
    q^{2}}}\left(1+{1\over 2q^{2}}\right)={3\sqrt{\pi}\over
  4}~{(1-\sigma)\Gamma(-\sigma)\over\Gamma({5\over 2}-\sigma)}\;,
\label{intrho}
\ee
(we have $\Re\sigma<0$ along the integration contour $c$).
As a result of all these manipulations we obtain the representation
\bea\label{E06}
E_0^{(1)}&=&
{\alpha\over 8\pi R}~\sum\limits_{nij}\left(X^{(1)}_{nij}+X^{(2)}_{nij}\right)
{1\over\Gamma(i+{1\over 2})}\nn\\
&&\times\int\limits_{c}{\d\sigma\over 2\pi\i}~{\Gamma({1\over 2}+i+\sigma)
\Gamma({\ep-1\over 2}+i+j+\sigma)\Gamma^2(-\sigma)(1-\sigma)
\over \Gamma({\ep\over 2}+i+j+\sigma)\Gamma({5 \over 2}-\sigma)}\nn\\
&&~~~~~~~~~~~~\times\left(4m^{2}R^{2}\right)^{\sigma}
\zeta_H\left(2n+2\sigma+\ep-2;{3\over 2}\right)\;.
\eea
With this representation of $E_0^{(1)}$ at our disposal, we are in a
position to separate the ultraviolet divergent contributions. They are 
identified with those poles of the integrand of \Ref{E06} that cross the 
integration contour when removing the regularization, i.e. in the limit
$\ep \rightarrow 0$. All such poles are contained in the addends $n=0,1$.
Namely, the zeta function possesses single poles at $\sigma=(3-\ep)/2$ $(n=0)$
and $\sigma=(1-\ep)/2$ $(n=1)$, respectively while the second Gamma function
in the numerator of \Ref{E06} contributes a simple pole at 
$\sigma=(1-\ep)/2$ when $n=0$. The addends with $n\ge 2$ only contain poles that
remain in the half--plane $\Re\sigma < 0$ when taking the limit 
$\ep\rightarrow 0$. The residues of the poles passing the integration contour
for $\ep\rightarrow 0$ give extra contributions to $E_0^{(1)}$.
These additional contributions are
\be 
E_0^{div,1}=-{16 \over9\pi}\alpha m^{3}R^{2}
\label{pol1}
\ee
from $\sigma=(3-\ep)/2$ and
\be E_0^{div,2}=-{4\over 15\pi} \alpha m
\label{pol2}
\ee
from $\sigma=(1-\ep)/2$. As we have anticipated, the ultraviolet divergent
contributions are proportional to $R^2$ and $R^0$. Hence their removal
can be interpreted as a redefinition of the parameters $\sigma$ and $k$
in the energy of the classical system \Ref{Eclass1}. The fact that the 
ultraviolet divergent contributions only appear as finite residues is
a peculiarity of the zeta--functional regularization.

The renormalized radiative correction to the ground state energy is now 
given by
\bea\label{Esub}
\mbox{$E_0^{(1)}$}^{\rm ren}&\equiv&E_0^{(1)}(\ep\rightarrow 0)-E_0^{div,1}
-E_0^{div,2}\\
&=&{\alpha\over 8\pi R}~\sum\limits_{n=0}^{\infty} 
\sum\limits_{ij}\left(X^{(1)}_{nij}+X^{(2)}_{nij}\right)
{1\over\Gamma(i+{1\over 2})}\nn\\
&&\times\int\limits_{c}{\d\sigma\over 2\pi\i}~{\Gamma({1\over 2}+i+\sigma)
\Gamma(-{1\over 2}+i+j+\sigma)\Gamma^2(-\sigma)(1-\sigma)
\over\Gamma(i+j+\sigma)\Gamma({5 \over 2}-\sigma)}\nn\\
&&~~~~~~~~~~~~\times\left(4m^{2}R^{2}\right)^{\sigma}\zeta_H\left(2n+2\sigma-2;
{3\over 2}\right)\;.\label{E0ren}
\eea
At the first glance $\mbox{$E_0^{(1)}$}^{\rm ren}$ \Ref{E0ren} looks just like 
$E_0^{(1)}$ \Ref{E06} with $\ep =0$. The difference is that the poles at 
$\sigma=(3-\ep)/2$ and $\sigma=(1-\ep)/2$ are now located on the right of 
the integration contour. The structure of the asymptotic series \Ref{E0ren} is 
well known. Shifting the integration contour to the left yields an expansion of 
$\mbox{$E_0^{(1)}$}^{\rm ren}$ in inverse powers of $mR$ (likewise, an expansion 
for $mR<< 1$ is obtained by moving it to the right).

Having separated the ultraviolet divergencies with the help of representation 
\Ref{E06} we return to eq. \Ref{E05}. When inserting $E_0^{(1)}$ \Ref{E05} 
into \Ref{Esub} the limit $\ep \to 0$ cannot directly be taken. Therefore we 
subtract the first two terms ($n=0,1$) of the asymptotic expansion \Ref{Bas}, 
\be 
B_{as}\equiv {s\over t}\sum\limits_{n=0}^{1}
\sum\limits_{i,j\atop n\le i,j\le 3n}
\left(X^{(1)}_{nij}+X^{(2)}_{nij}\right)
{s^{2i}t^{2j}\over \nu^{2n}}\; ,
\label{As}
\ee
from the integrand and add them again. In this way we split 
$\mbox{$E_0^{(1)}$}^{\rm ren}$ according to 
\be 
\mbox{$E_0^{(1)}$}^{\rm ren}= E_{f}+E_{as}
\label{E0ren2}
\ee
into 
\be 
E_{f}={i\over 2\pi R} \sum\limits_{l=1}^{\infty}\int\limits_{-\infty}^{\infty}
{\d k \over 2\pi}~{\nu \over \left(k^{2}+\nu^{2}\right)^{\frac{\ep}{2}}}
 \int\limits_{4m^2R^2}^{\infty}{\d  q^{2}\over q^{2}}~
{\rm disc}\tilde{\Pi}
\left(\frac{q^2}{R^2}\right)~(B_{1}+B_{2}-B_{as})
\label{Ef}
\ee
and
\be 
E_{as}={i\over 2\pi R} \sum\limits_{l=1}^{\infty}\left[\int\limits_{-\infty}^{\infty}
{\d k\over 2\pi}~{\nu \over \left(k^{2}+\nu^{2}\right)^{\frac{\ep}{2}}}
 \int\limits_{4m^2R^2}^{\infty}{\d  q^{2}\over q^{2}}~
{\rm disc}\tilde{\Pi}
\left(\frac{q^2}{R^2}\right)~B_{as}\right]
 ~-E_0^{div,1}-E_0^{div,2}\;.
\label{Eas}
\ee
In the first contribution, $E_{f}$, the limit $\ep \rightarrow 0$ can be carried
out under the signs of the integral and the sum because they converge now. 
This form is suited for a numerical evaluation, if necessary. 
In the second contribution, $E_{as}$ we pass again to the Mellin-Barnes 
representation and observe that $E_{as}$ is identical with the first two 
addends ($n=0,1$) of the asymptotic series in \Ref{E0ren}. 

Representation \Ref{E0ren2} is the final result for the renormalized
radiative correction of order $\alpha$ to the ground state energy of a 
conducting sphere.

Next we turn to the interesting physical situation where the radius $R$
of the sphere is large compared to the Compton wavelength of the electron, 
i.e. $mR>>1$. In other words, we are interested in the leading order term
when expanding $\mbox{$E_0^{(1)}$}^{\rm ren}$ in inverse powers of $mR$. 
The representation \Ref{E0ren} serves us as a starting point. As already 
mentioned above, the expansion in inverse powers of $mR$ is achieved by shifting 
the integration contour to the left. The leading term of the desired expansion 
is given by the residue at $\sigma =-1/2$. A table
of all addends contributing a pole at $\sigma =-1/2$ is displayed in 
Appendix B. We observe that the addends with $n=0,1,2$ contain also double 
poles which lead to the appearance of logarithmic contributions 
whereas the terms with $n>2$ show only simple poles. It is
therefore technically convenient to calculate the contributions with
$n\le 2$ and $n>2$ separately. Accordingly, we split the leading order
term into two pieces, 
$E_0^{(1)}=E_{n\le 2}+E_{n>2}+O(m^{-2}R^{-3})$.
The residues contributing to $E_{n\le 2}$ are calculated
by standard methods and we find
\bea 
E_{n\le 2}&=&-{\alpha\over 420\pi} 
{\log(mR) \over mR^2}-{\alpha\over 8\pi mR^2}\left\{\frac{1447573}{16934400}
+\frac{31}{5376}C\right.\nn\\
&&\left.+\frac{393}{4480}\log 2-\frac{3}{4}\zeta_H'(-3;\frac{3}{2})
+\frac{3}{16}\zeta_H'(-1;\frac{3}{2})\right\}\nn\\
&=&-{\alpha\over mR^{2}}~\{7.5788\cdot 10^{-4}\log(mR)+6.4735\cdot 10^{-3}\}.
\label{Eas3}
\eea
The contribution $E_{n>2}$ can only be calculated
numerically. For this purpose, we cast it into the more convenient form
\be\label{En2}
E_{n>2}=\frac{3\alpha}{32\pi m R^2}
\sum_{l\ge 1}\nu^2\int\limits_{0}^{\infty}\d k\left\{
B_1^{0}+B_2^{0}-\sum_{n=0}^{2}\sum_j\left(
X^{(1)}_{n0j}+X^{(2)}_{n0j}\right){t^{2j-1}\over \nu^{2n}}\right\},
\ee
where we have introduced the notation
\be 
B_{1}^{0}={1\over 2\nu}{1\over I_{\nu}(k)K_{\nu}(k)}\ee
and 
\be B_{2}^{0}=
{t^{-2}\left(1-{t^{2}\over 4\nu^{2}}\right)/\left( 1-{1\over 4\nu^{2}}\right)
\over 2\nu I_{\nu}(k)K_{\nu}(k)+k^{2}\left({1\over \nu+\frac{1}{2}}
I_{\nu+1}(k)K_{\nu+1}(k)+{1\over \nu-\frac{1}{2}}
I_{\nu-1}(k)K_{\nu-1}(k) \right)}\;.
\ee
Expression \Ref{En2} can also directly be obtained from \Ref{Ef}. For this 
purpose, the term with $n=2$ of the asymptotic expansion \Ref{Bas} has to be 
subtracted from the integrand in addition to $B_{as}$. Then, the leading term 
of the expansion of the integrand in inverse powers of $q^2$ yields $E_{n>2}$. 

The remaining task is the numerical evaluation of $E_{n>2}$.
The result is
\be 
E_{n>2}= -{\alpha\over mR^{2}}~9.8230\cdot 10^{-6}\; ,
\label{Efnum}
\ee
where contributions up to $l=50$ have been taken into account (the sum over
$l$ converges as $\sum_{l}\nu^{-3}$, $(\nu=l+1/2)$).

So we finally obtain the radiative correction in leading order $1/(mR)$
\be 
E_0^{(1)}=-{\alpha\over  mR^{2}}~
\left(7.5788\cdot 10^{-4}\log mR +6.4833\cdot  10^{-3}\right)+
O\left(\frac{1}{m^2R^3}\right)\;.
\label{E0as}
\ee

\section{Summary}

In the present paper we have calculated the radiative correction
to the ground state energy of a perfectly conducting spherical shell. 
Being proportional to the fine structure constant $\alpha$, it is small 
compared to the Casimir energy itself. In a realistic physical situation 
where the radius $R$ of the sphere is much larger than the Compton
wavelenght $\lambda_c$ of the electron, the radiative correction is
proportional to the ratio $\lambda_c/R$. In contrast to the case of two 
parallel conducting planes, the radiative correction for a sphere 
exhibits a logarithmic dependence on $\lambda_c/R$.

The calculations have been performed in the framework of general
covariant perturbation theory with the boundary conditions incorporated
as constraints. A brief review of quantization with boundary conditions 
in this formalism has been supplied. As a valuable advantage of this
approach, the ghost degrees of freedom are not affected by the boundary 
conditions so that they need not be taken into account. The use of the 
formalism has been demonstrated by recalculating the Casimir energy 
including radiative correction for two perfectly conducting planes.
The Casimir energy of a conducting sphere could be reproduced as well.
As expected from the known heat kernel expansion the removal of the 
geometry--dependent ultraviolet divergencies can be absorbed in a 
redefinition of the parameters \Ref{Eclass1} of the external system
represented by the surface ${\cal S}$. In the calculation of the vacuum 
graph to order $\alpha$ the electron mass $m$ and the charge $e$ need 
not be renormalized. 

A thorough investigation of the renormalization of gauge theories in the 
presence of boundary conditions beyond the vacuum structure as well as 
the analogous calculations within the MIT bag model seems to be an 
interesting future perspective. 

The authors thank K. Kirsten and D. Vassilevich for helpful discussions. 

\begin{appendix}
\section{The coefficients $X^{(1,2)}_{nij}$}
The coefficients $X^{(1,2)}_{nij}$ in formula \Ref{Bas} are:
$$ X^{(1,2)}_{000}=1\;, $$

$$  X^{(1)}_{1ij}=\left({\begin{array}{cccc} 
0&-{1\over 8}&{3\over 4}&-{5\over 8}\\[5pt]
 {1\over 8}&0&0&0\\[5pt]
   -{3\over 4}&0&0&0\\[5pt]
 {5\over 8}&0&0&0
\end{array}}\right)\;,\qquad
X^{(2)}_{1ij}=\left({\begin{array}{cccc} 
  0&-{1\over 8}&-{3\over 4}&{7\over 8}\\[5pt]
 {1\over 8}&0&0&0\\[5pt]
{3\over 4}&-{3\over 2}&0&0\\[5pt]
{5\over 8}&0&0&0  \end{array}}\right)\;,
 $$
\vspace{4pt}
$$ X^{(1)}_{2ij}=\left({\begin{array}{ccccccc} 
 0&0&-{{25}\over {128}}&{{139}\over {32}}&-{{1039}\over {64}}&
    {{663}\over {32}}&-{{1105}\over {128}}\\[5pt]
 0&-{1\over {64}}&{3\over {32}}&-{5\over {64}}&0&0&0\\[5pt]
{{27}\over {128}}&{3\over {32}}&-{9\over {16}}&{{15}\over {32}}&0&0&0\\[5pt]
 -{{145}\over {32}}&-{5\over {64}}&{{15}\over {32}}&
    -{{25}\over {64}}&0&0&0\\[5pt]
 {{1085}\over {64}}&0&0&0&0&0&0\\[5pt]
-{{693}\over {32}}&0&0&0&0&0&0\\[5pt]
 {{1155}\over {128}}&0&0&0&0&0&0     \end{array}}\right)
$$
and
 $$ X^{(2)}_{2ij}=\left({\begin{array}{ccccccc} 
 0&0&{{23}\over {128}}&-{{143}\over {32}}&{{1181}\over {64}}&
    -{{819}\over {32}}&{{1463}\over {128}}\\[5pt]
 0&-{1\over {64}}&-{3\over {32}}&{7\over {64}}&0&0&0\\[5pt]
 -{{21}\over {128}}&{{21}\over {32}}&-{3\over 4}&{{57}\over {32}}&
    -{{21}\over {16}}&0&0\\[5pt]
 {{125}\over {32}}&-{{545}\over {64}}&
    -{{15}\over {32}}&{{35}\over {64}}&0&0&0\\[5pt]
-{{595}\over {64}}&{{105}\over 4}&0&0&0&0&0\\[5pt]
-{{63}\over {32}}&-{{315}\over {16}}&0&0&0&0&0\\[5pt]
{{1155}\over {128}}&0&0&0&0&0&0 
\end{array}}\right)\;. $$

\section{The pole structure of $\mbox{$E_0^{(1)}$}^{ren}$ in \Ref{E0ren}}

Here we list all addends of \Ref{E0ren} which contribute a pole
at $\sigma=-\frac{1}{2}$ when shifting the integration path to the 
left. They result from the zeta function and the first two Gamma
functions in the numerator of \Ref{E0ren}: 
$$
\begin{array}{ccccccc}
n&i&j&\Gamma(i) & \Gamma(i+j-1) &\zeta_H(2n-3;3/2) \\ \hline
0&0&0&\times&\times&&(\mbox{double pole})\\
1&0&1&\times&\times&&(\mbox{double pole})\\
1&0&2&\times&&&\\
1&0&3&\times\\
1&1&0&&\times\\
2&0&2\dots 6&\times&&\times&(\mbox{double poles})\\
2&2&0\dots 4&&&\times\\
2&4&0\dots 2&&&\times\\
2&6&0&&&\times\\
3\dots\infty& 0 & n\dots 3n &\times &&&$\phantom{{double poles}dd}$.\\
\end{array}
$$
The residues can be calculated by standard methods. The double poles yield 
the logarithm $\log mR$.
\end{appendix}

\end{document}